
\tolerance=10000
\input phyzzx

\font\mybb=msbm10 at 12pt
\def\bbbb#1{\hbox{\mybb#1}}
\def\Z {\bbbb{Z}}
\def\R {\bbbb{R}}

\def \g {g_S}
\def \m {m_S}
\def\gym {g_{YM}}
\def\gef {g_{eff}}
\def \aa {\alpha}
\def \bb {\beta}
\def \gg {\gamma}

\def \kk {\kappa}

\def \th {\theta}

\def \sss {\Sigma}

\def\sym{super Yang-Mills}

\def \ti {\tilde}

\def \2 {{1 \over 2}}
\def \3 {{1 \over 3}}
\def \4 {{1 \over 4}}
\def \5 {{1 \over 5}}
\def \6 {{1 \over 6}}
\def \7 {{1 \over 7}}
\def \8 {{1 \over 8}}
\def \9 {{1 \over 9}}
\def \00{{ \infty }}

\def\gy {g_{YM} }

\def \qq {\qquad}

 \def\unit{\hbox to 3.3pt{\hskip1.3pt \vrule height 7pt width .4pt \hskip.7pt
\vrule height 7.85pt width .4pt \kern-2.4pt
\hrulefill \kern-3pt
\raise 4pt\hbox{\char'40}}}




 \REF\Mat{T. Banks, W. Fischler, S. Shenker, and L. Susskind,
hep-th/9610043, Phys. Rev.
{\bf D55} (1997) 5112.}%
\REF\kp{  M. Douglas, D.  Kabat, P. Pouliot and  S. Shenker,
Nucl. Phys. {\bf B485} (1997) 85,
 hep-th/9608024.}
\REF\brev{For a review see T. Banks,
hep-th/9706168, hep-th/9710231.}%
\REF\Sus{L. Susskind, hep-th/9704080.}
\REF\taylor{W. Taylor, hep-th/9611042, Phys. Lett. {\bf B394} (1997) 283.}%
\REF\motl{L. Motl, hep-th/9701025.}
\REF\IIAbs{T. Banks and N. Seiberg,hep-th/9702187.}
\REF\IIAVV{R. Dijkgraaf,E. Verlinde and H. Verlinde, hep-th/9703030.}
\REF\dvv{R. Dijkgraaf, E. Verlinde, and H. Verlinde,  hep-th/9603126;
 hep-th/9604055; hep-th/9703030; hep-th/9704018.}
\REF\rozali{M. Rozali, hep-th/9702136.}
\REF\brs{M. Berkooz, M. Rozali, and N. Seiberg,
hep-th/9704089}
\REF\fhrs{W. Fischler, E. Halyo, A. Rajaraman and L. Susskind,
hep-th/9703102.}%
\REF\MatTor{N. Seiberg,
hep-th/9705221.}%
 \REF\seinotes{  N. Seiberg,
hep-th/9705117.}
\REF\mem{A. Losev, G. Moore and S.L. Shatashvili, hep-th/9707250; I. Brunner
and
A. Karch,
hep-th/9707259, hep-th/9801154; A. Hanany and G. Lifschytz, hep-th/9708037; R.
Argurio and L.
Houart, hep-th/9710027.}
\REF\Sen{A. Sen, hep-th/9709220.}
\REF\Seib{N. Seiberg,
hep-th/9710009.}%
 \REF\malstro{J. Maldacena and A. Strominger,   hep-th/9710014.}
\REF\mal{J. Maldacena,   hep-th/9711200.}
 \REF\itmal{
N. Itzhaki,  J. Maldacena, J.  Sonnenschein  and S. Yankielowicz,
hep-th/9802042.}
\REF\witt{E. Witten, hep-th/9802150.}
\REF\hum{C.M. Hull,    hep-th/9710165, hep-th/9712075.}
\REF\humat{C.M. Hull, hep-th/9711179.}
\REF\hooft{G. 't Hooft, Nucl. Phys. {\bf B72} (1974) 461; Nucl. Phys. {\bf B75}
(1974) 461.}
\REF\HT{C.M. Hull and P.K. Townsend, hep-th/9410167.}
\REF\tay{W. Taylor, hep-th/9705116.}
\REF\hor{ G. Horowitz and A. Strominger,
  Nucl. Phys {\bf B360} (1991) 197.}
\REF\huten{C.M. Hull, Phys.Lett. {\bf B357 } (1995) 545, hep-th/9506194.}
\REF\GravDu{C.M. Hull, Nucl.Phys. {\bf B509 } (1998) 216, hep-th/9705162.}
\REF\rub{P. Ruback, Commun. Math. Phys. {\bf 107} (1986) 93.}
\REF\kkmo{E. Bergshoeff, B. Janssen and T. Ortin,
hep-th/9706117.}
\REF\HS{C.M. Hull and B. Spence, Nucl. Phys. {\bf B353} (1991) 379.}
\REF\berg{E. Bergshoeff, E. Eyras  and  Y. Lozano, hep-th/9802199.}
\REF\kur{R. Khuri and T. Ortin, Phys. Lett. {\bf B373} (1996) 56.}
\REF\shein{H.J. Scheinblatt, hep-th/9705054.}
\REF\kes{E. Keski-Vakkuri and P. Kraus, hep-th/9706196.}
\REF\perr{J. Pierre, hep-th/9707102.}
\REF\bran{A. Brandhuber, N. Itzhaki, J. Sonnerschein and S.
Yankielowicz, hep-th/9711010.}
\REF\dar{A. Ahar and G. Mandal, hep-th/9803004.}
\REF\mai{P. Dobiasch and D. Maison, Gen. Rel. Gravitation {\bf 14} (1982) 231.}
\REF\giba{A. Chodos and S. Detweiler,  Gen. Rel. Gravitation {\bf 14} (1982)
879;
D. Pollard, J. Phys. {\bf A16} (1983) 565.}
\REF\gib{G.W. Gibbons and D.L. Wiltshire, Ann. Phys. {\bf 167} (1987)
201; Erratum Ann. Phys. {\bf  176} (1987) 393.}
\REF\lar{V. Balasubramanian, R. Gopakumar and F. Larsen,
 hep-th/9712077. }
\REF\mar{N. Marcus and A. Sagnotti, Phys. Lett. {\bf 135B} (1984) 85; Nucl.
Phys. {\bf B256} (1985) 77.}
\REF\dix{Z. Bern, L. Dixon, D. Cunbar, M. Perelstein and J. Rozowsky,
hep-th/9802162.}
\REF\hs{P.S. Howe and K.S.
Stelle, Phys. Lett. {\bf 137B} (1984) 175.}
\REF\pari{G. Parisi, Nucl. Phys. {\bf B100} (1975) 368 and {\it On
Non-Renormalizable Interactions}, in \lq New Developments in Quantum
Theory and Statistical Mechanics', eds. Levy and Mitter (Plenum Press,
1977).}

\Pubnum{ \vbox{  \hbox {QMW-98-09} \hbox{NSF-ITP-98-024}\hbox{hep-th/9803124}}
}
\pubtype{}
\date{March, 1998}

\titlepage

\title {\bf  Decoupling Limits in M-Theory }

\author{C.M. Hull}
\address{Physics Department, Queen Mary and Westfield College,
\break Mile End Road, London E1 4NS, U.K.}
\andaddress{Institute for Theoretical Physics, UCSB, Santa Barbara, CA 93106,
USA}

\vskip 0.5cm

\abstract {Limits of a system of $N$ D$n$-branes in which the bulk and string
degrees of freedom decouple to leave a \lq matter' theory are investigated and,
for $n>4$, either
give a free theory or require taking $N\to \infty $. The decoupled matter
theory is described at low
energies by  the $N\to \infty $ limit of $n+1$ dimensional \sym, and at high
energies by a free type
II string  theory in a curved space-time. Metastable  bound states of D6-branes
with mass $M$ and
D0-branes with mass $m$ are shown to have an energy proportional to
$M^{1/3}m^{2/3}$ and decouple,
whereas in matrix theory they only decouple in the large $N$ limit. }

\endpage

\chapter {Introduction}

The study of field theory limits of M-theory or string theory in the presence
of
branes has proved very fruitful.
In particular, the limit used in matrix theory   has been related to the
infinite momentum frame or discrete light cone quantization of M-theory
[\Mat-\Seib].
Similar limits have led to new non-trivial \lq matter' theories in
$5+1$ dimensions [\dvv-\seinotes] arising from limits of M-theory in which
gravity decouples.
However, the construction of matrix theories has proved problematic for
M-theory
compactified on an $n$-torus if $n>5$ [\Sen,\Seib] (and perhaps for $n=5$
[\malstro]), and is
related to the difficulty in constructing non-trivial theories in greater than
6
dimensions. Here we will   consider a different limit from that considered in
[\Mat-\Seib],   which appears to give  rise to a decoupled matter theory in 5+1
or 6+1
dimensions,
but which does not seem to be directly related to any kind of light-cone
quantization of M-theory. This limit involves  taking $N \to \infty  $ in the
underlying $U(N)$ gauge theory and is closely related to the limits
  considered by Maldacena et al  [\mal,\itmal].

Consider weakly-coupled type II string theory in the presence of $N$ coincident
D$n$-branes. (We will mostly restrict ourselves to $n<7$, due to the problems
with having arbitrary numbers of D$n$-branes, for $n\ge 7$.)
This is described by $n+1$ dimensional  super Yang-Mills  with $U(N)$ gauge
symmetry and
coupling constant $\gym$, coupled to 10-dimensional supergravity with
gravitational coupling $\kk$,
together with stringy corrections characterised by the string mass scale
$\m=1/\sqrt{\alpha ' }$.
The coupling of the bulk degrees of freedom to the fields on the brane are also
characterised by $\kk$.
At strong coupling, the type IIA theory (with $n$ even) is described by
M-theory
and the type IIB theory (with $n$ odd) by the S-dual type IIB theory.
The field theory limits that it is natural to consider [\kp] are ones in which
$  \kk \to 0$ so that gravity and the bulk degrees of freedom decouple and $\m
\to \infty $ so that the string excitations decouple, while $\gym$ remains
fixed, in
order to obtain a non-trivial theory with gauge interactions.
Such a limit is possible for $n<5$, and is central to the construction of the
matrix theory for M-theory compactified on $T^n$. (In this construction, the
type II
theory is taken to be compactified on the T-dual torus, and the D$n$-branes are
wrapped on this torus.)
However, the coupling constants are not independent and such a limit is not
possible for $n\ge 5$.
In particular, for $n=5$, the string mass, after the S-duality
transformation, is $\hat \m = \gym ^{-1}$, so that taking $\hat \m \to \infty $
and
decoupling the string modes is possible only if $\gym \to 0$. Instead, in
[\brs,\MatTor],
$\hat \m = \gym ^{-1}$ was kept fixed,   resulting in a 5+1 dimensional
string theory [\brs,\MatTor]. This is the world-volume theory of $N$ NS
5-branes
and the bulk degerees of freedom do not completely decouple from the throat
region of the 5-brane background
[\malstro].
For $n=6$, the 11-dimensional Planck mass is $m_p=\gym ^{-2/3}$, so that
if gravity is to decouple, again it is necessary to take $\gym \to 0$. In
[\mem,\Sen,\Seib],
$m_p=\gym ^{-2/3}$ was kept finite, so that gravity does not decouple and the
result is M-theory on an ALE $A_{N-1}$ singularity. (For $n=7$, one
would have $\gym
^{-2/3}=m_p^{(10)}$, where $m_p^{(10)}$ is the 10-dimensional Planck mass, and
gravity would again not decouple.)

In this limit for $n=6$, the radius $R$ of the   11'th dimension in the
M-theory
description  diverges so that the theory decompactifies, and at the same time
the BPS states of the bulk theory include   D0 branes, which have masses of
order $1/R$, and so
become massless. This decompactification and the presence of an infinite number
of massless states are further obstacles to the formulation of a matrix theory
for
M-theory on $T^6$ [\Sen,\Seib].

In this paper, an alternative limit will be considered, in which for $n=5,6$
$\gym \to 0$ so that bulk and string modes decouple.
If $N$ is kept fixed, the result is a free theory, which is consistent but not
interesting.
In order to obtain a non-trivial theory, $N$ will be taken to infinity at the
same time as $\gym \to 0$, while keeping the effective coupling
$$ g_N = \gym \sqrt N
\eqn\abc$$
fixed, giving the 't Hooft limit [\hooft] of the super Yang-Mills  theory, in
which only the
planar
Feynman diagrams survive. In this limit, for large but finite $N$, the bulk
and/or string modes do not decouple and are presumably crucial to the
consistency of the theory, but in the infinite $N$ limit, the bulk and string
modes all appear to decouple, and presumably leave a consistent matter theory
in
5+1 or
6+1 dimensions. In addition, $R$ will be kept fixed in the limit, so that the
11'th dimension does not decompactify and there are no extra massless states.
Whereas  in  the usual matrix limit, the
characteristic scale of the D0-branes on the dual torus, which is the
11-dimensional Planck scale of the T-dual theory, plays a key role,
  the limit considered here for $n=6$  corresponds to keeping
the states with energies characterised by the energy scale of the D0-branes on
$T^6$ that
are becoming light. This can be thought of as performing a Weyl rescaling at
the
same time as $m_S$ is scaled, in such a way that $R$ is kept constant and the
mass of
the D0-branes is kept fixed.

Limits in which $R$ is not kept fixed, but scales with $N$ so that $R\to 0$ as
$N\to \infty $ will also
be considered. For $n=4$, the D4-brane is described by an M5-brane
wrapped on a circle of radius $R$. The world-volume theory of $N$ M5-branes is
the $U(N)$ self-dual
tensor multiplet theory in 5+1 dimensions, which dimensionally reduces to 4+1
super Yang-Mills  with coupling
constant
$\gym ^2 =R$. Taking the 't Hooft limit of the theory corresponds to taking $R
\propto 1/N$, so
that as $N
\to \infty $, the extra dimension disappears, leaving the large $N$ limit of
4+1
\sym.
For $n=6$, taking $R\to 0$ completely decouples the D0-branes from the bulk
theory, leaving 6+1
\sym.

 The large $N$ behaviour has recently been considered in [\mal,\itmal] in the
context of the usual
limit
[\kp]. The limit considered here is closely related, as
will be described in section 4.
For energy scales $U$ that are
small compared to the scale set by the effective coupling $g_N^2=\gym^2N$, the
theory
is described by the large $N$ limit of \sym, while for energies large compared
to the effective coupling, the description is in terms of string theory in a
particular background, as in [\mal,\itmal].

\chapter {The Limits of Matrix Theory}

 Consider the system of  coincident  $N$ D$n$-branes   in IIA
string theory (for $n$ even) or in type IIB string theory ($n$ odd),
 with
string tension $m_S^2$ and string  coupling $g_S$.
At low energies and weak coupling, the bulk degrees of freedom are described
by
type II supergravity and the brane dynamics by super Yang-Mills (with 16
supersymmetries)  in $n+1$ dimensions, with gauge group $U(N)$. The kinetic
terms include
$$
\m^8 \int d^{10}x \sqrt{-g} e^{-2\phi}R
+\m^{n-3} \int d^{n+1}\sigma \sqrt{-g} e^{-\phi}F^2
$$
giving the following effective gravitational coupling $\kappa$ and
10-dimensional
Planck
scale $m_p^{(10)}$:
$$m_p^{(10)}\equiv \kappa ^{-1/4}
= m_s \g ^{-1/4}
\eqn\abc$$
while the Yang-Mills coupling $g_{YM}$ is
$$
g_{YM}^2= { g_S
\over  m_S^{n-3}
}\eqn\abc$$
Here
$$ \g= e^{\phi _ \infty }
\eqn\abc$$
where $\phi _ \infty $ is the asymptotic value of the dilaton $\phi$.

For $n$ even, the type IIA theory arises  as
M-theory compactified
on $S^1$ with radius $R$, with
$$
R={g_S\over m_S}, \qquad
m_p= {m_S \over g_S^{1/3}}
\eqn\abc$$
where $m_p$ is the 11-dimensional Planck mass, and the type IIA description  is
 appropriate only when $\g $ is small.
For $n$ odd, an S-duality transformation  takes the type IIB theory   to
a dual type IIB theory with    string mass
scale $\hat m_S$   and
string coupling $\hat g_S$ given by
$$\eqalign{&
 \hat m_S =   m_S   g_S^{-1/2}
\cr
&
  \hat g_S =   g_S^{-1}
\cr}
\eqn\abc$$
so that the strong coupling limit $\g \to \infty $ is described in the dual
theory
as the weak coupling limit $\hat \g \to 0$.

The BPS spectrum is most easily considered if the theory is compactified on
an
$n$-torus $T^n$
with radii $R_i$, with the $N$ D$n$-branes wrapped on the torus.
Consider the BPS states of the bulk type II theory compactified on $T^n$.
There are momentum states with mass of order $1/R_i$ and  string winding
states
with mass of order $\m^2 R_i$.
There are Dirichlet  $p$-branes wrapped on a $p$-cycle for $p\le n$.
Such a $p$-brane,
wrapped on the first $p$ cycles of $T^n$ has mass of order:
$$g_S^{-1} m_S^{p+1} \Big(\prod_{i=1}^p R_i\Big) \eqn\pmass$$
For $n\ge 5$ there are
NS 5-branes wrapped on five of the circular dimensions, with
mass
 $$
g_S^{-2} m_S^6 \Big(\prod_{i=1}^5 R_i\Big) \eqn\abc$$
In addition, for $n\ge 6$ there are states arising from wrapping Kaluza-Klein
monopoles of the type II theory, which should be included as they are T-dual to
the NS 5-branes. These depend on 6 radii $R_i$, $i=1,....,6$, one
of which ($R_6$, say) is distinguished as the fibre of the Kaluza-Klein
monopole
solution. These
   have mass
 $$
g_S^{-2} m_S^8 \Big(\prod_{i=1}^5 R_i\Big) R_6^2 \eqn\abc$$
Some  of these BPS states an form bound states with the D$n$-branes, in which
case the binding
energy can alter   the  mass formulae; this will be discussed in section 6.

If the type II theory is put on an $n$-torus with radii $R_i$
and the $N$ D$n$-branes are wrapped on $T^n$, then a T-duality
  transformation takes this to
$N$ D0-branes in the  type IIA theory with string coupling $\ti \g$ and string
mass
$\ti \m$ on the dual torus $\ti T^n$ with radii $\ti R_i$. The parameters  $\ti
g_S$, $\ti m_S$
and $\ti R_i$
are given by
$$\eqalign{&
\ti \m =\m , \qquad \ti R_i =\m^{-2} R_i^{-1}
\cr
&
\ti \g= \g / \prod_{i=1}^n (\m R_i)
\cr}
\eqn\abc$$
This is in turn related to an $\ti M$-theory compactified on a space-like
circle, and boosting and rescaling and taking a certain limit gives M-theory
compactified on a null circle (regarded as the limit of a boosted space-like
circle)
and hence the matrix theory representation  of M-theory [\Seib].

A natural decoupling limit to consider is one in which $m_S \to \infty $ and
$g_{YM}$ is kept fixed, and, if the theory is  on a torus $T^n$, the radii
$R_i$ are also kept fixed.
In the T-dual picture, this corresponds to
taking the limit $\ti \g \to 0$ of the T-dual theory, keeping fixed
$$
\ti m_p \ti R_i  =\ti g_S^{-1/3} \ti \m \ti R_i , \qquad
\ti m _p ^2 \ti R=\ti  m_S \ti g_S^{1/3}
\eqn\mlim$$
and this is the limit used in the  matrix theory  construction   [\Mat-\Seib].
  We will refer here to
  the limit   $m_S \to \infty $ with $g_{YM}$   fixed as the matrix
limit; it gives a
theory related to  $n+1$ dimensional
super Yang-Mills (on $T^n\times \R$).
Then
$$\eqalign{&
\g=\hat \g ^{-1}=\gy ^2 \m^{n-3}, \qq m_p^{(10)}=\gy^{-1/2}\m^{(7-n)/4},
\cr
&
\hat m_s=\gy^{-1 }\m^{(5-n)/2},
\qq R=\gy ^2 \m ^{n-4}, \qq m_p= \gy^{-2/3}\m^{(6-n)/3}
\cr}
\eqn\abc$$
If the D$n$-branes are wrapped on a torus and the  radii $R_i$ are kept fixed
in
the limit, the BPS states surviving in the limit are  the momentum modes and
wrapped D$p$-branes with $p\le n-4$, since the D$p$-brane mass \pmass\ can be
rewritten
as
$$
 \gym^{-2} \m^{p-n+4} \prod_{i}^{p}{R_i}
\eqn\abc$$
For $n<3$, $\g \to 0$, so that the D-brane picture with weakly coupled strings
is appropriate,
$\m \to \infty  $ so that the
massive
string modes decouple and
$m_p^{(10)}\to \infty $, so that supergravity and the bulk degrees of freedom
decouple, leaving $n+1$ dimensional super Yang-Mills.
For $n=3$, $\g=g_{YM}^2$ remains finite, but the string and bulk degrees of
freedom decouple
and the limiting theory is   four-dimensional super Yang-Mills, with $N=4$
superconformal symmetry.
For $n>3$, $\g \to \infty $ and we need to use the dual strong coupling theory.

For  even $n>3$, we use M-theory with Planck mass $m_p \sim \m ^{(6-n)/3}$ and
 the radius $R$ of the 11'th dimension
$R\sim \m ^{n-4}$, so that for $n=4$, $m_p \to \infty $ and gravity and the
bulk
degrees of freedom decouple, together with the membrane excitations that reduce
to string
excitations for small $R$, while
$R$  remains finite. The M-theory description of the D4-brane is as an M5-brane
wrapped around the
11'th dimension, so that the D4-brane wrapped on $T^4$ corresponds to an
M5-brane wrapped on $T^5$.
The limiting theory for $n=4$  is the M5-brane world-volume theory, which is
a
generalisation of the $(2,0)$ supersymmetric self-dual  tensor theory with
$U(N)$ gauge symmetry [\rozali-\seinotes].
The D0-branes have mass proportional to
$1/R=\gym ^{-2}$ and arise from solitons in the  super Yang-Mills  theory
obtained by
lifting instantons in 4 Euclidean dimensions
to $4+1$ dimensions.
They are interpreted as Kaluza-Klein modes from the point of view of the 5+1
dimensional theory. The D0 branes with mass $1/R$  combine with the momentum
modes with mass $1/R_i$ (all of which form bound states at threshold with the
D4-brane) to give the
momentum modes on $T^5$ in the M-theory description, transforming as a {\bf 5}
of  $SL(5,\Z)$, which is the U-duality
group for M-theory compactified on $T^4$ [\HT]\foot{The
representation of $SL(5,R)$ on $R^5$ induces
a representation of $SL(5,Z)$ on the 5-dimensional lattice $Z ^5$, which we
will refer to as the {\bf 5} representation of $SL(5,Z)$, and similarly for
other U-duality groups.}.

For $n=6$, the Planck scale  $m_p$ remains finite, but $R\to \infty $.
A single D6-brane corresponds to a Kaluza-Klein monopole of M-theory,
and
the solution with $N$ D6-branes lifts to a multi-Kaluza-Klein monopole
solution of M-theory, in the singular limit in which $N$ monopoles coincide.
The matrix limit gives   the   $R \to \infty $ limit of this, which is
  an $A_{N-1}$ ALE singularity, so we obtain M-theory on an ALE singularity
with gravity and the bulk degrees of freedom not decoupling [\Sen,\Seib].
The BPS states considered above whose  mass remains  finite and non-zero   in
this limit are the
6 momentum modes, the 15 wrapped D2-branes and the 6 wrapped NS 5-branes; each
forms a bound state
at threshold with the D6-brane. These combine   to form a {\bf 27} of
$E_6(\Z)$,
which is the
U-duality group  for M-theory compactified on $T^6$. In addition, the bulk
D0-branes
are becoming massless in the limit, leading to an infinite number of massless
states; these   correspond to gravitons (and their
superpartners) moving in the 11'th dimension, which is decompactifying in the
limit [\Sen]. In section 6, the metastable bound state formed between the D0
and D6-branes will be
considered; these have finite energy for fixed $N$ and so play a role in the
world-volume theory,
but decouple in the large $N$ limit of
[\Mat].

For $n=5$, the $N$ D5-branes become $N$ Neveu-Schwarz 5-branes in the S-dual
type IIB theory, which is free in the limit  as $\hat \g \to 0$,
with $m_p^{(10)}\to \infty $ and gravity and the bulk decoupling, but
the string scale $\hat m _S$ remaining finite.
For $n=5$, the marginally bound wrapped D1-strings   in
the D5-brane have finite mass and become fundamental strings moving in the NS
5-brane after the S-duality; the 5 momentum modes and 5 string winding modes
combine to form a {\bf 10} of
$SO(5,5;\Z)$.
This limit has been argued to give rise to a 5+1  dimensional \lq non-critical'
string
theory with string tension
 $\hat m _S^2$ and (1,1) supersymmetry, whose zero slope limit is  (1,1)
super Yang-Mills  [\MatTor,\seinotes].
For the NS 5-brane of the type IIA theory,
it was argued in [\malstro] that in regimes in which   string theory in
a
near-extremal 5-brane space-time is a good description,  there is Hawking
radiation with temperature $T_H \propto \hat \m$, so that it appears that
there
is not a complete decoupling from the bulk in the throat region of the NS
5-brane space-time.
The NS 5-brane of the type IIB theory is the same supergravity background as
for
the type IIA case (the RR gauge fields all vanish) and the arguments of
[\malstro]
apply in this case also, so that  there appears not to be a complete
decoupling from the bulk in the throat region.

\chapter {Other Limits}

For the usual matrix limit with $n=6$, there are two problems: $m _p $ remains
finite, so that gravity doesn't decouple, and $R$ becomes infinite, so that the
bulk D0-branes become massless.
For $n=5$, $\hat \m $ remains finite, so that one obtains a string theory
instead of a field theory, and it appears that
there is not a complete decoupling from the bulk, due to Hawking radiation in
the throat region.
This leads naturally to the search for more general limits in which $m_p \to
\infty $
while $R$ remains finite for $n=6$, and $\hat \m \to \infty $ for $n=5$. Such
limits could lead to new  decoupled \lq matter' theories in 6+1 or 5+1
dimensions,
but are unlikely to be related to light cone quantizations of M-theory.

In the T-dual description in terms of ${\ti {D0}}$-branes on $\ti T^n$, one can
consider
replacing the usual limit $\ti \g \to 0$ keeping \mlim\ fixed with the limit
$\ti \g \to 0$  keeping
$$
 g_S^{-A} \ti \m \ti R_i , \qquad
 \ti  m_S \ti g_S^{B}
\eqn\abc$$
fixed for some constants $A,B$ (generalising the usual limit \mlim\ in which
$A=B=1/3$). The new limits that will be proposed
correspond to choosing
  $$A= B={1 \over n-1}
\eqn\abc$$
for $n>1$, so that for $n=4$, the limit considered here is the same as the
usual
one.
Whereas the usual limit [\Mat,\kp] focuses on scales of order the Planck scale
$\ti m_p$,
the new limit for $n=6$ focuses on the scale  associated with the branes
that are becoming massless (the D0-branes on $T^6$, or the ${\ti {D6}}
$-branes wrapped on $\ti T^6$ in the T-dual description). This can be thought
of
as performing a $\ti \g $-dependent  Weyl-rescaling at the same time as scaling
$\ti \g$ in such a way that $R$ and the mass of the bulk $D0$-branes is kept
fixed.
It will be more convenient to present the results in terms of the theory on
$T^n$, however.

For $n=6$, in order
for the bulk modes to decouple, we clearly need a  limit in which  $m_p\to
\infty
$,
but for $n=6$, $\gym^2=m_p^{-3}$, so that
this would require $\gym \to 0$. Similarly, for a decoupled theory with
$n=5$,
it seems a limit in which $\hat \m \to \infty $ is needed, but for $n=5$, $\gym
=\hat \m^{-1}$, so that again this would require $\gym \to 0$.
In either case, such a limit leads to decoupling of bulk modes, but
if $N$ is kept fixed, the result is a free super Yang-Mills  theory. If, on the
other
hand,
$N$ is taken to infinity at the same time, while keeping
$$g_N ^2\equiv \gym^2N
\eqn\abc$$
fixed, then the result is   the 't Hooft limit of
$n+1$ dimensional \sym, for $n=5,6$.
If in addition a rescaling is chosen so that $R$ is kept fixed in the limit,
then
the bulk D0-branes will remain of finite non-zero mass in the limit.
We will subsequently consider more general limits in which $R$ is not kept
fixed, and which
give rise for $n=4$ to the 't Hooft limit of 4+1 \sym.

We now propose a limit satisfying both of these
criteria, and which generalises  to all $n$, although the meaning is unclear
for
$n>6$.
In this limit we take $\m \to \infty $ and keep fixed $R \equiv \g /\m$ both
for
even and for odd $n$;
for even $n$, $R$ is the radius of the 11'th dimension.
Then
$$\eqalign{
& \g =1/\hat \g =
\m R, \qq m_p^{(10)}=\m^{3/4} R^{-1/4},\qq
m_p =\m^{2/3} R^{-1/3},
\cr
&
\hat m_s=\m^{1/2} R^{-1/2}, \qq \gym ^2 =R \m ^{4-n}
\cr}
\eqn\abc$$
Thus in the limit $\m \to \infty $, the mass scales $\m ,
m_p^{(10)},
m_p,
\hat m_s$ all diverge, so that bulk modes
and
string modes decouple for all $n$.
The string coupling $\g \to \infty  $ in this limit, so that the dual strong
coupling description is appropriate. For the type IIA theory ($n$ even), this
gives M-theory with $m_p \to \infty $ and a {\it finite } size $R$ of the
compact
11'th dimension,
while for type IIB ($n$ odd) this gives the weak coupling limit $\hat \g \to 0$
of the S-dual type IIB theory, with $\hat \m , \hat m _p^{(10)} \to \infty $.

When on a torus $T^n$, whose radii $R_i$ are kept fixed in the limit, most of
the  bulk
BPS
states considered above decouple.
The masses of the   winding strings, the wrapped D$p$-branes for $p>0$ and the
wrapped NS
5-branes
all diverge  and these BPS states are absent in the limit.
The momentum modes and the D0 branes  (which are present for $n$ even) have
finite mass, and the mass of the
D0-branes is now finite and given by
$1/R$ (for all even $n$), whereas in the matrix limit the D0-branes became
massless
for $n=6$.

The Yang-Mills coupling constant diverges in the limit for $n<4$, is finite
for
$n=4$ and tends to zero for $n>4$.
For $n=4$, the limit is exactly the same as the usual matrix limit.
For $n=3$, the type IIB S-duality leads to a super Yang-Mills  S-duality
transforming $\gym
\to
\hat \gym \equiv 1/\gym$,
and $\hat \gym \to 0$ in the limit.
For $n=3$ and $n>4$, taking this limit with $N$ fixed gives a free theory,
while taking the limit with $N\to \infty $ and
$g_N=\gym ^2 N$ fixed gives a non-trivial limit.

For $n>4$, keeping $g_N$ and $R$ fixed implies
$$
\m = \aa N^{1\over n-4}, \qq \aa = \left(  g_N^2 \over R \right) ^{1\over n-4}
\eqn\abc$$
and for $n=3$, keeping $\hat g_N ^2 = \hat \gym ^2 N=N/\gym^2$ fixed
implies
$$
\m = \bb N
, \qq \bb = {1\over R \hat g_N^2}
\eqn\abc$$
giving the rate at which $\m$ diverges with $N$.

For all $n$ these
limits are ones in which
in which bulk and string modes decouple, and which has a
description
in terms of
 super Yang-Mills  in $n+1$ dimensions.
If $N$ is kept fixed, the result is a free theory, while if $N$ is taken to
$\infty $ for $n \ge 3$, we
obtain an 't Hooft limit.
For $n=3,5$, we obtain  in this way the large $N$ limit  of $n+1$ \sym. For
$n=6$, we obtain $6+1$
\sym, either as a free theory or a large $N$ limit,  together with bulk $D0$
branes with mass of
order
$1/R$. If we now take
$R
\to 0$, these D0-branes decouple leaving pure super Yang-Mills  on the brane.
For finite $R$, there could be extra degrees of freedom corresponding to the
D0-branes in the 6+1
dimensional theory, but these could be consistently decoupled by taking $R\to
0$; this will be
discussed further in sections 5,6.

For $n=4$ with fixed $N$, we obtain the usual limit consisting of the
$(2,0)$
tensor theory in $5+1 $ dimensions with $U(N)$ symmetry, on a circle of radius
$R$. The dimensional
reduction to $4+1$ dimensions gives super Yang-Mills  with $\gym ^2 =R$; this
is
non-renormalizable, and the
extra degrees of freedom needed at short distances are provided by the
Kalua-Klein modes.
If we take $N\to \infty $ in this theory while keeping
$\gym ^2 N$ fixed, then $R\sim 1/N \to 0$, and the resulting theory is the
large
$N$ limit of
4+1 \sym, with the extra dimension disappearing.

Although consistent perturbative quantum  super Yang-Mills  theories   do not
exist in 5,6
or
7
dimensions with finite gauge groups, we have seen that the large $N$ limit of
5,6
or 7 dimensional $U(N)$ theories
 arise as limits of M-theory in which the bulk modes appear to decouple, and so
could be consistent
matter theories in these dimensions.
We will return to this in section 7.

Note that in these limits for the theory on $T^n$, only the
geometric
$SL(n,\Z)$ subgroup of
the full U-duality group will   survive,
as the Kaluza-Klein modes for $n=4$, the string winding modes for $n=5$ and the
membrane wrapping
modes and NS 5-brane  wrapping modes for $n=6$ all decouple.
(Note that for $n=6$, a limit in which  the NS 5-branes survived and the
membranes did not could
lead to a theory with duality group
$SL(6;\Z)\times SL(2;\Z)$, which is a maximal subgroup
of $E_6(\Z)$.
 This
  $SL(2,\Z)$ S-duality identified in [\humat] would interchange  the six
5-branes with the six
momentum modes, and acts on $\gym$ and the generalised $\th $ angle of
[\humat],
which   combine
into a complexified coupling constant.)

\chapter{The Supergravity Solutions}

The type II solution for $N$ coincident D$n$-branes is, for $n<7$,
[\hor,\itmal,\huten]
$$
\eqalign{& ds^2=f_n^{-1/2}(-dt^2+dx_1^2+...+dx_n^2)+f_n^{1/2}
(dx_{n+1}^2+...+dx_9^2),
\cr &
e^
\phi=e^{\phi_{\infty}}
 f_n^{-(n-3)/4},
\cr &
 A_{0...n}=-\2 (f_n^{-1}-1) ,
\cr}
\eqn\abc$$
where $f_n$ is a harmonic function of the transverse coordinates
$x_{n+1},...,x_9$, which, for  $N$ coincident branes at the origin $r=0$
(where the radial coordinate $r=
(x_{n+1}^2+...+x_9^2)^{1/2}$)
is
$$
 f_n= 1+ \left( {a \over r } \right) ^{7-n}
 \eqn\abc$$
 where
 $$ a^{7-n}= {\g N d_n \over \m ^{7-n}}
 ={R N d_n \over \m ^{6-n}}=
 {\gg  ^2  \m ^{2(n-5)}}
 \eqn\abc$$
 Here $d_n$ is a constant given in [\itmal], and
 $$
 \gg =\sqrt{d_n}g_N=\gym\sqrt{d_n N}
 \eqn\abc$$
 In the asymptotic region $r/a >> 1$, the solution approaches
 flat space with $\phi \to \phi _ \infty $, giving string theory in flat space
with
coupling constant $\g$, while in the near region $r/a <<1$, it is
appproximately
described by the   geometry
$$   ds^2=
\left( {a \over r } \right) ^{-(7-n)/2}
dx^2_{||}+
\left( {a \over r } \right) ^{(7-n)/2}(dr ^2 + r^2 d\Omega^2_{8-n})
\eqn\met$$
and is   singular as $ r \to 0$ (for $n \ne 3$).
The dilaton is given by
$$\phi = \phi _{\infty } + {(7-n) (n-3)\over 4} \log  {(r/a)}
\eqn\abc$$ so that
as $r\to 0$, $ e^\phi \to 0 $ for $3<n<7 $ and $ e^\phi \to \infty  $ for
$n<3$.

In the limits considered here, if $N$ is kept fixed then $a \to 0 $ for $n<6$,
and is finite ($a=d_6 RN$) for $n=6$, while if $N \to \infty  $ keeping $\gym
^2 N$
fixed, $a \to \infty  $. If $r$ is rescaled at the same time, the limiting form
will
be the asymptotic flat geometry if $a/r$ remains small, but will be the near
geometry \met\ if $a/r$ becomes large.
The limiting forms of these geometries in the matrix limit have been considered
in [\mal,\itmal], in the limit in which the radial coordinate is also scaled so
that
$$U = \m ^2 r
\eqn\abc$$
is held fixed; this corresponds to keeping fixed
the energy of a string joining D-branes separated by a distance $r$ in the
D-brane
description, and $U$ is the energy scale of the super Yang-Mills  description
[\mal,\itmal].
In terms of $U$,
$$
 f_n= 1+ \left( {b \over U } \right) ^{7-n}
 \eqn\abc$$
 where
 $$ b^{7-n}= {\g N d_n   \m ^{7-n}}
 ={R N d_n   \m ^{8-n}}=
 {\gg  ^2  \m ^{4}}
 \eqn\abc$$

Then in any  limit in which $\m \to \infty $ while $U$ and $g_N=\gym\sqrt{  N}$
are kept fixed,
 the solution becomes [\itmal]
$$\eqalign{& ds^2=\m ^{-2}\left(
 {{U^{(7-n)/2}}\over {\gg }}dx^2_{||}+
 {{\gg}\over {U^{(7-n)/2}}}
 dU^2+\gg U^{(n-3)/2}
d\Omega^2_{8-n}
\right) ,
\cr & e^{\phi}=
\gym^2\left(  {g_N^2  }\over {U^{7-n}}\right) ^{ {3-n}\over {4}} \sim
 { {\gef^{(7-n)/2}}\over {N}}
  \cr }
\eqn\thro$$
where  $g_{eff}^2=g_N^2U^{n-3}$ is the effective coupling at energy scale $U$.
This is the geometry of the near region \met,
 rescaled by a factor $\m ^2$ which diverges in the limit.
 It is the remarkable fact that the metric only depends on
 $\gym $ and $N $ in the combination $\gym ^2 N$ that leads to the structure of
the large $N$ theory [\itmal].

In particular, if $\gym \to 0, N \to \infty $, as in the case here, we have the
metric \thro\ but
$e^\phi \to 0$.
Note that in this limit, although $\g= e^{\phi _\infty } \to \infty $ for
$n>3$,
$e^\phi = e^{\phi _\infty } f_n^{-(n-3)/4}
\to 0$.
In terms of $\gef$ the curvature {\cal R}
associated with the metric   is [\itmal]
$$
\m ^{-2}{\cal  R} \approx {1\over \gef } \sim \sqrt{{ U^{3-n} \over \gym^2 N }}
\eqn\abc$$
This will be small if $g_{eff}$ is large, and this is a necessary condition for
the supergravity solution to be reliable.
As   the position-dependent string coupling
$e^\phi$ becomes
zero in this limit,  the weakly-coupled type II picture is reliable.
Then if $g_{eff}>>1$, i.e. at energy scales $U$ such that
$$U^{n-3} >> g_N ^ 2
\eqn\abc$$
the limiting theory is described by free type II superstrings in the
near-geometry background \thro, while for energies
$$U^{n-3} << g_N ^ 2
\eqn\abc$$
the effective coupling $g_{eff}$ is small, and the theory is described by the
large $N$ limit of super Yang-Mills  for $n=5,6$.

\chapter {The Role of the D0-branes  for $n=6$.}

We now return to the role of the $D0$-branes for $n=6$.
For $n=6$, the free limit or large $N$ limit of super Yang-Mills  was obtained
by taking
$R\to 0$
after the limits
described above. If $R$ is kept finite, there are D0-branes of finite mass
surviving in the bulk and
which might lead to extra
 degrees of freedom in the world-volume theory.
The interaction between D0 branes and D6 branes is not completely understood,
but  D0-branes do not
form a stable BPS bound state with the D6-branes, and there is a repulsive
force. If a D0-brane is
put on a D6-brane, it would tend to shrink to zero size and then escape from
the
brane into the
bulk. However, there are non-supersymmetric bound states that are quadratically
stable [\tay], so
that D0-brane states might be expected to play a role in the world-volume
theory
for finite $R$,
and this need not be superymmetric. In the free theory arising from the limit
$\m \to \infty $ keeping
$N$  fixed, the D0-brane
could correspond to an extra free scalar moving in a compact dimension,
corresponding to the fibre of
the multi-Taub-NUT space.

In [\GravDu], it was argued that the supersymetric world-volume dynamics of the
M-theory Kaluza-Klein monopole $\R^{6,1}\times (Taub-NUT)$ should be described
by the super Yang-Mills  multiplet
in 6+1 dimensions. In a covariant approach, the 6+1 world-volume theory should
include a vector field
and 11 scalars
$X^M$, of which $6+1$ scalars can be identified with the world-volume
coordinates by choosing a
static gauge, leaving 4 scalars taking values in the Taub-NUT space. The vector
multiplet in 7
dimensions has only 3 scalars however, so that we obtain a vector multiplet and
an extra scalar,
taking values in the  fibre of the Taub-NUT, which is a circle of radius $R$.
Translations in this fibre direction do not constitute true deformations, and
are pure gauge as
they constitute an isometry of the space-time; in particular, there is no
modulus associated with
them [\rub]. In [\kkmo], it was proposed that the effective description of the
4
scalars should be a
sigma-model whose target space is the Taub-NUT space, in which the fibre
isometry
is gauged, leaving a theory of 3 scalars, which fit into the vector multiplet.
In [\kkmo], a kinetic term was proposed for the bosonic degrees of freedom, and
shown to
have appropriate properties, and in particular the reduction to 5+1 dimensions
gives the D6-brane
action.
The general form of the Wess-Zumino term for    gauged sigma-models in any
dimension was given in
[\HS], and it is straightforward to add  such a term to the kinetic term of
[\kkmo] to obtain an
action that reduces to the full D6-brane action; Wess-Zumino terms for such
actions have also been
considered in [\berg].

The D0-brane degree of freedom clearly corresponds to the scalar taking values
in the Taub-NUT
fibre, and its zero-mode drops out because of the
isometry symmetry of the target space. The scalar can be decoupled from the
Taub-NUT sigma-model
by gauging, as in [\kkmo], or by taking
$R\to 0$. It cannot fit into a 7-dimensional supermultiplet with the vector and
other scalars.

To discuss whether the D0-brane survives in the  limit $N\to \infty $, we need
to
know how the energy
of the metastable D0-D6 bound state depends on $N$ and $\m$.
In the next section, we will show that, at least in regimes in which
supergravity is reliable, the
bound state of $k$ D0-branes to the $N$ D6-branes does not decouple for matrix
theory with fixed $N$,
but does in the $N\to \infty $ limit of [\Mat]. For the limits proposed here
with $\m \to \infty $ and
$R,R_i$ fixed, these decouple   both for the free limit with $N$ fixed and in
the  large $N$ limit.

\chapter {D-Brane Bound States}

The    D$n$-D$n-p$ brane bound state has been extensively studied for $p=2,4$.
For $p=2$, there is a true bound state preserving half the supersymmetries, and
the D$n$-D$n-2$ brane
bound state wrapped on an
$n$-torus gives a state of mass
$M$ satisfying
$$M^2=Q^2+P^2
\eqn\abc$$
For $p=4$, there is a marginally boud state preseserving $1/4$ of the
supersymmetry, and the wrapped
D$n$-D$n-4$ brane bound state has mass
$$M=|Q|+|P|
\eqn\abc$$
Here
$P$ is the mass of $N$ D$n$-branes,
$$P=g_S^{-1} m_S^{n+1} \Big(\prod_{i=1}^n R_i\Big) \eqn\pmass$$
and $Q$ is the mass of $k$ D$n-p$ branes
$$Q=kg_S^{-1} m_S^{n-p+1} \Big(\prod_{i=1}^{n-p} R_i\Big) \eqn\pmass$$
The   energy $E=M-|P|$ is then $E=|Q|$  for the marginally bound states with
$p=4$, while for
$p=2$,
$$E\sim {Q^2\over 2|P|}
\eqn\abc$$
if $Q/P <<1$, as will be the case for large $N$, so that $E\sim 1/N$.
The formulae are also the mass formulae for the black hole solution obtained by
dimensionally
reducing the brane solutions to $10-n$ dimensions (or less).
The black hole has
   two   electric charges $Q,P$ with respect to two different gauge fields and
preserves $1/2$
($p=2$) or $1/4$ ($p=4$) of the supersymmetry.

For $p=6$,  the D6-D0 brane configuration is only metastable and is not
supersymmetric.
Its dimensional reduction
should give a dyonic  black hole in 4 dimensions carrying an electric charge
$Q$ and a magnetic
charge
$P$ with respect to the same gauge field, which is the vector field obtained
from the dimensional
reduction of the Ramond-Ramond one-form.
Such dyonic black holes have been considered in this context in
  [\kur-\dar]. The relevant black hole solution is that of [\mai,\giba,\gib],
and, for the extreme black hole, the mass $M$ satisfies
$$M^2 =4 \left( Q^2+P^2 \right)-\sss ^2
\eqn\mas$$
where
$\sss$ is the dilaton charge defined by
$$
{1 \over 6}\sss = {Q^2\over \sss + \sqrt{ 3} M}+{P^2\over \sss - \sqrt{ 3} M}
\eqn\mart$$
 For the magnetically charged solution with $Q=0$,
$$P=M, \qq  \sss =-
\sqrt 3M, \qq Q=0
\eqn\abc$$
while
For the electrically charged solution with $P=0$,
$$Q=M, \qq  \sss =
\sqrt 3M, \qq P=0
\eqn\abc$$

The dilaton charge can be eliminated from these formulae (using e.g. the
parameterisation of [\mai]) to give
$$M=\left( Q^{2/3}+P^{2/3} \right) ^{3/2}
\eqn\abc$$
Then if $|Q|<<|P|$,
$$ E=M-P
\sim {3\over 2} P^{1/3}Q^{2/3}
\eqn\abc$$
plus corrections of order
$|Q|^{4/3}|P|^{-1/3}$

Defining the volume of the 6-torus in string units,
$$\hat V_6=m_S^{6}V_6 =m_S^{6} \Big(\prod_{i=1}^6 R_i\Big)
\eqn\abc$$
the energy can be written as
$$\eqalign{
E&\sim {3\over 2} N^{1/3}k^{2/3}{1\over R} \hat V_6^{1/3}
\cr &
= {3\over 2} N^{1/3}k^{2/3}{1\over \gym ^2}
V_6^{1/3}
\cr &
= {3\over 2} N^{4/3}k^{2/3}{1\over g_N ^2}
V_6^{1/3}
\cr}
\eqn\eng$$
This formula will apply whenever
$Q << P$;
as $|Q/P| \sim k/(N\hat V_6) $, we have that $|Q/P| \to 0$  if $\m \to \infty $
with
$R_i$ fixed, as is the case for the matrix and decoupling limits
discussed here. Then
the higher order corrections, which are suppressed by a factor of
$|Q/P|^{2/3}$,
vanish in the limit  and
the formula  \eng\ is exact.
 In the usual matrix limit in which $\gym,N,V_6$ are kept fixed, this gives
$E \propto N^{1/3} m_p^3$ and this remains finite, so that the finite $N$
matrix theory should
include these bound states, but if $N$ is taken to infinity (keeping $\gym$
fixed, as in [\Mat,\kp])
these states decouple.
For the limits proposed here, $  V_6 $ is kept fixed
and $\gym \to 0$, so that these states decouple either if $N$ is kept fixed or
if
$N\to \infty $, $g_N$ fixed, in which case
$E\sim N^{4/3}$.
If   $R$ is kept fixed, the states in which D0-branes bind to the D6-branes
decouple, even though
the bulk D0-branes remain at finite mass, while if $R\to 0$, the bulk D0-branes
decouple also.
 (Note that if $R$ and the string-scale volume $\hat V_6$ were kept fixed,
these states would remain for finite $N$ but decouple as $N\to \infty $, with
$E\sim N^{1/3}$.)

\chapter {Discussion}

We have seen that for $n=3,4,5,6$, there is a limit of the theory of $N$
D$n$-branes that gives
the $N \to \infty $ limit of $n+1$ dimensional \sym, for which the Planck mass
and
string mass both
tend to infinity, so that the bulk and stringy degrees of freedom should all
decouple. For $n=4$,
this involves taking $R\to 0$, so that the extra dimension disappears,
while for
$n=3,5$ this involves
$\hat
\g
\to 0,\hat
\m
\to \infty $, so that  the stringy degrees of freedom decouple.
 The D$n$-brane action has higher-derivative terms such as those from the
expansion of the
Dirac-Born-Infeld kinetic term, but these are suppressed by powers of the
string
scale and
disappear in the limit. If there is a complete decoupling of all other degrees
of freedom, then
this limit should give a consistent quantum theory, and the simplest way in
which this might
occur would be if the large $N$ limits of these super Yang-Mills  theories were
finite.
As the theories with finite $N$   are not finite or renormalizable
for $n>3$, this would be quite
remarkable.

If one considers the situation with $N$ large but finite, the theory should
make
sense (assuming
M-theory does) and   there are extra degrees of freedom that become important
at
the short
distance scale    $l_{YM}= \gym ^{2/(n-3)}$, with $\gym$ small but finite.
For $n=4$, these are Kaluza-Klein modes asscociated with a circle of radius
$R=l_{YM}$, for $n=5$
these include string excitations of non-critical strings with string mass $\hat
\m =1/l_{YM}$, and
for $n=6$ these include bulk modes coupling at the Planck scale $m_p=
1/l_{YM}$.
However, taking $N\to \infty $ takes the effective cut-off $l_{YM} \to 0$, and
it
appears that the extra
short-distance degrees of freedom decouple.
Some evidence for this is given in [\lar], where it is shown that for certain
processes and in
certain kinematical regimes, certain non-planar    super Yang-Mills  processes
correspond to
supergravity effects
that are suppressed by powers of the Planck scale.

For finite $N$, the divergence structure of super Yang-Mills  in $D=n+1$
dimensions
has been explicitly calculated to two loops by Marcus and Sagnotti [\mar], and
their results have
been confirmed in [\dix], while the implications of non-renormalisation
theorems
were investigated in
[\hs]. For $D=8,10$, there is a divergence occuring at one-loop, for $D=7,9$
there are divergences at
2-loops, calculated in [\mar]. For $D=5,6$, the theory is finite to two-loop
order [\mar], but there are  supersymmetric counterterms at 3-loops for $D=6$
and at 4-loops for
$D=5$ that are allowed by $N=2$ non-renormalisation theorems [\hs].
In [\dix], it was argued that for $D=5$,
the 4-loop and 5-loop $D=5$ divergences should be
absent also, and that  the first  ultra-violet divergence should occur at
6-loops (or higher).
Whenever counterterms are not forbidden, they are expected to actually occur.

The 2-loop counterterm in 7 dimensions is of the form
$C_{abcd}F^aF^bDF^cDF^d+....$ (with certain contractions of the  space-time
indices, suppressed here)
and there are two
possible group-theory factors    $C_{abcd} $, one of which is associated with
planar graphs and the
other with non-planar graphs [\mar,\dix].  The actual counterterm is a linear
combination of the two,
so that in the large $N$ limit the non-planar one  is suppressed, but the
planar
one survives.
Thus for $D=7$, the theory defined by taking the large $N$ limit while keeping
the ultra-violet
cut-off fixed remains ultra-violet divergent, i.e. there are still planar terms
which
diverge as the
cut-off is taken to infinity.
For $D=5,6$ the allowed counterterms can also have planar or non-planar
group-theory factors, and
there is no reason to expect that the coefficients of the planar ones should
vanish, although it
would be interesting to have explicit confirmation of this.

The likely persistance of ultra-violet divergences in the large $N$ limit at
first seems
disappointing. It could mean that there is not a complete decoupling in this
limit, but it is hard to see how this could be the case, since
the constants $1/\m , 1/m_p$ etc governing these couplings have all gone to
zero. Alternatively,
it has been proposed that  the large $N$ limit can be used to give a finite
formulation
of perturbatively non-renormalisable
theories (there is an extensive literature, but see, for example, [\pari]), and
something similar may
play a role here.
However, finiteness would have meant that large $N$ super Yang-Mills  provides
a good
description at all
energy scales, which would be hard to reconcile with the proposed high energy
beahviour of
[\mal,\itmal] and section 4.
The picture that seems to fit best is that the large $N$ limit of the theory of
$N$
D-branes proposed here gives a limit in which the bulk and string modes
decouple, leaving a  theory that is consistent (since it is derived from
M-theory), is
described at low energies by the 't Hooft limit of  \sym, and   the extra
degrees of freedom that
become relevant at short distances are
described by
the curved space string theories proposed
in [\mal,\itmal],
as we shall now argue.

Consider the case of $n=4$, which is the one that is best understood.
For finite $R=\gym ^2$ and finite $N$, the theory is described by the (2,0)
U(N)
tensor theory in 6
dimensions, compactified on a circle of
radius
$R$. This is believed to be a consistent finite superconformally invariant
field
theory in 6
dimensions [\rozali-\seinotes]. The large $N$ limit has been argued to exist,
and to be described at
high energies by the type IIA string theory on  the limiting form of the
D4-brane background
[\itmal].
 Taking $R\to 0$ while keeping $N$ fixed, the Kaluza-Klein modes decouple,
leaving free
super Yang-Mills  theory in 4+1 dimensions, with $\gym \to 0$.
Taking the 't Hooft limit of the $D=6$ theory, keeping $NR=N\gym ^2$ fixed,
should give a
consistent theory. As $U(N)$ super Yang-Mills  is expected to be a good
effective
description at length scales
that are large compared to $R$, the $N \to \infty , R\to 0$ limit might be
expected
to correspond to
the large $N$ limit of $4+1$ \sym. However, the large $N$ limit of the
(2,0) theory should still be finite, while the large $N$ limit of the $4+1$
super Yang-Mills  may still be
divergent. If so, this would  suggest  that
the two theories are not the same, and that the large $N$ super Yang-Mills  is
only an
effective description.
The only scale surviving in the limit is that set by $g_N^2=RN$, which is kept
fixed in the limit,
so it seems reasonable to assume that this is the effective cut-off.
The resulting theory is  effectively described by super Yang-Mills  at length
scales large
compared with
$g_N^2$, but extra degrees of freedom must become important at distances small
compared to this.
However, at distances small compared to  $RN$, we can use the description
proposed in [\itmal]
and the extra degrees of freedom that enter are provided   by   type IIA
string
theory
on the limiting form of the D4-brane background. (For large enough $N$, the
other regime of [\itmal]
described by M-theory on an M5-brane background is absent.)

This suggests  the following picture. The $D=6$ (2,0) theory gives a finite
theory for all $N,R$.
If $R>0$ and $N$ is small, Kaluza-Klein modes are important at length scales
small compared to $R$
and provide the extra degrees of freedom needed at short distances in the
non-renormalisable $D=5$
\sym. In the 't Hooft limit, $R \to 0$ so that the Kaluza-Klein modes decouple,
but it appears that
the resulting $D=5$ theory (the 't Hooft limit of $D=5$ \sym) remains
non-renormalisable.
The effective cut-off must be $RN=g_N^2$, and the extra degrees of freedom
needed to regulate the
theory at shorter distance scales are provided by IIA string theory in the
near-region of a D4-brane
background.

A similar picture should then apply for the non-renormalisable super Yang-Mills
 theories in
$D=6,7$; for finite
$N,\gym$, the extra degrees of freedom needed at distances small compared with
$l_{YM}$ are string
modes in
$D=6$ and certain bulk modes in $D=7$.   For the 't Hooft limit in which
$l_{YM}
\to 0$, the
resulting theory is again presumably not finite, even though the usual
effective cut-off $l_{YM} $
has been removed, and
the only surviving scale in the theory is that set by $g_N$, which should be
the
effective cut-off.
 The
planar limit of the
super Yang-Mills  is a good effective description at distance scales large
compared with
this, and the type II
string theory in the near region limit of the D$n$-brane background
is a good effective description at distance scales small compared with
the $g_N$ scale.
This generalises the duality between    $D=4$  super Yang-Mills  and type IIB
string theory in anti-de Sitter space
[\mal,\witt]: at weak coupling
$g_{eff}<<1$ (where
$g_{eff}^2=Ng_{YM}^2 U^{n-3} $ is the dimensionless effective coupling at
energy
$U$) there is a
super Yang-Mills  description, while for $g_{eff} >>1$ there is a description
in terms of
free type II string theory in the near
region limit of a D-brane background.

To summarise, the theory of $N$ D$n$-branes in the
large $N$ limit described here gives a theory
in which gravity and the bulk degrees of freedom decouple, along with the
string
excitations, to
leave a \lq matter theory'. If M-theory is consistent, these theories should be
also.
For $n\le 4$, such a decoupling can be obtained for finite $N$, but for $n>4$,
one either obtains a
free theory or one must
take $N\to \infty  $.
 For weak effective coupling $g_{eff}$ the theory is the large $N$ limit
of $U(N)$ \sym, while for strong coupling $g_{eff}>>1$ it is described by the
free type II string
theory in the near region limit of the D$n$-brane space-time proposed in
[\mal,\itmal]. For $n=3$,
the
super Yang-Mills  theory is perturbatively finite, and the strong coupling
formulation  in
terms of
free type IIB strings in  anti-de Sitter space and the weak coupling
formulation
 in terms of large
$N$
super Yang-Mills  are dual formulations of the same theory [\mal,\witt]. For
$n>3$, there
are again two dual
formulations, but here the energy scale $U$ plays a role, so that the super
Yang-Mills
decription gives the
behaviour  at energies low compared to
$U_N=g_N^{-2/(n-3)}$ while the high energy behaviour is that of
the free string theories of [\mal,\itmal].
There are thus two scales that enter into the D-brane theory: $\gym$ and $g_N$,
and
extra short distance degrees of freedom enter at whichever scale is the larger.
 In the limits in
which $N$ is kept fixed, extra degrees of freedom enter at the energy scale set
by $\gym$ which are
Kaluza-Klein modes for $n=4$, string degrees of freedom for $n=5$ and bulk
degrees of freedom for
$n=6$. In the large $N$ limits considered here in which $\gym \to 0$, these
degrees of freedom
associated with $\gym$ decouple, but the new degrees of freedom that enter at
the energy scale set
by $g_N$ are associated with free string theories in D-brane space-times.

\ack
{I would like to thank Tom Banks, Mike Douglas, Juan Maldacena, Steve Shenker
and  Paul Townsend for helpful discussions. This work was partly supported by
the NSF, Grant No. PHY94-07194, and partly by the EPSRC. }


\refout

\bye